\PassOptionsToPackage{unicode}{hyperref}
\PassOptionsToPackage{hyphens}{url}
\PassOptionsToPackage{dvipsnames,svgnames,x11names}{xcolor}
\documentclass[
]{article}

\usepackage{amsmath,amssymb}
\usepackage{iftex}
\ifPDFTeX
  \usepackage[T1]{fontenc}
  \usepackage[utf8]{inputenc}
  \usepackage{textcomp} 
\else 
  \usepackage{unicode-math}
  \defaultfontfeatures{Scale=MatchLowercase}
  \defaultfontfeatures[\rmfamily]{Ligatures=TeX,Scale=1}
\fi
\usepackage{lmodern}
\ifPDFTeX\else  
\fi
\IfFileExists{upquote.sty}{\usepackage{upquote}}{}
\IfFileExists{microtype.sty}{
  \usepackage[]{microtype}
  \UseMicrotypeSet[protrusion]{basicmath} 
}{}
\makeatletter
\@ifundefined{KOMAClassName}{
  \IfFileExists{parskip.sty}{%
    \usepackage{parskip}
  }{
    \setlength{\parindent}{0pt}
    \setlength{\parskip}{6pt plus 2pt minus 1pt}}
}{
  \KOMAoptions{parskip=half}}
\makeatother
\usepackage{xcolor}
\ifx\paragraph\undefined\else
  \let\oldparagraph\paragraph
  \renewcommand{\paragraph}[1]{\oldparagraph{#1}\mbox{}}
\fi
\ifx\subparagraph\undefined\else
  \let\oldsubparagraph\subparagraph
  \renewcommand{\subparagraph}[1]{\oldsubparagraph{#1}\mbox{}}
\fi

\usepackage{longtable,booktabs,array}
\usepackage{calc} 
\usepackage{etoolbox}
\makeatletter
\patchcmd\longtable{\par}{\if@noskipsec\mbox{}\fi\par}{}{}
\makeatother
\IfFileExists{footnotehyper.sty}{\usepackage{footnotehyper}}{\usepackage{footnote}}
\makesavenoteenv{longtable}
\usepackage{graphicx}
\makeatletter
\def\maxwidth{\ifdim\Gin@nat@width>\linewidth\linewidth\else\Gin@nat@width\fi}
\def\maxheight{\ifdim\Gin@nat@height>\textheight\textheight\else\Gin@nat@height\fi}
\makeatother
\setkeys{Gin}{width=\maxwidth,height=\maxheight,keepaspectratio}
\makeatletter
\def\fps@figure{htbp}
\makeatother
\newlength{\cslhangindent}
\newlength{\csllabelwidth}
\newlength{\cslentryspacingunit} 
\newenvironment{CSLReferences}[2] 
 {
  \setlength{\parindent}{0pt}
  \ifodd #1
  \let\oldpar\par
  \def\par{\hangindent=\cslhangindent\oldpar}
  \fi
  \setlength{\parskip}{#2\cslentryspacingunit}
 }%
 {}
\usepackage{calc}

\newcommand{\CSLLeftMargin}[1]{\parbox[t]{\csllabelwidth}{#1}}
\newcommand{\CSLRightInline}[1]{\parbox[t]{\linewidth - \csllabelwidth}{#1}\break}

\usepackage{arxiv}
\usepackage{orcidlink}
\usepackage{amsmath}
\usepackage[T1]{fontenc}
\makeatletter
\@ifpackageloaded{caption}{}{\usepackage{caption}}
\AtBeginDocument{%
\ifdefined\contentsname
  \renewcommand*\contentsname{Table of contents}
\else
  \newcommand\contentsname{Table of contents}
\fi
\ifdefined\listfigurename
  \renewcommand*\listfigurename{List of Figures}
\else
  \newcommand\listfigurename{List of Figures}
\fi
\ifdefined\listtablename
  \renewcommand*\listtablename{List of Tables}
\else
  \newcommand\listtablename{List of Tables}
\fi
\ifdefined\figurename
  \renewcommand*\figurename{Figure}
\else
  \newcommand\figurename{Figure}
\fi
\ifdefined\tablename
  \renewcommand*\tablename{Table}
\else
  \newcommand\tablename{Table}
\fi
}
\@ifpackageloaded{float}{}{\usepackage{float}}
\floatstyle{ruled}
\@ifundefined{c@chapter}{\newfloat{codelisting}{h}{lop}}{\newfloat{codelisting}{h}{lop}[chapter]}
\floatname{codelisting}{Listing}

\makeatother
\makeatletter
\@ifpackageloaded{caption}{}{\usepackage{caption}}
\@ifpackageloaded{subcaption}{}{\usepackage{subcaption}}
\makeatother
\makeatletter
\@ifpackageloaded{tcolorbox}{}{\usepackage[skins,breakable]{tcolorbox}}
\makeatother
\makeatletter
\@ifundefined{shadecolor}{\definecolor{shadecolor}{rgb}{.97, .97, .97}}
\makeatother
\ifLuaTeX
  \usepackage{selnolig}  
\fi
\IfFileExists{bookmark.sty}{\usepackage{bookmark}}{\usepackage{hyperref}}
\IfFileExists{xurl.sty}{\usepackage{xurl}}{} 
\hypersetup{
  pdftitle={Revealing Patterns of Symptomatology in Parkinson's Disease: A Latent Space Analysis with 3D Convolutional Autoencoders},
  pdfauthor={E. Delgado de las Heras; F.J. Martinez-Murcia; I.A. Illán; C. Jiménez-Mesa; D. Castillo-Barnes; J. Ramírez; J.M. Górriz},
  colorlinks=true,
  linkcolor={blue},
  filecolor={Maroon},
  citecolor={Blue},
  urlcolor={Blue},
  pdfcreator={LaTeX via pandoc}}

\newcommand{\runninghead}{A Preprint }
\renewcommand{\runninghead}{Revealing Patterns of Symptomatology in
Parkinson's Disease }
\title{Revealing Patterns of Symptomatology in Parkinson's Disease: A
Latent Space Analysis with 3D Convolutional Autoencoders}
\author{
\textbf{E. Delgado de las Heras}\\\\Dpt. Signal Theory, Networking and
Communications, University of Granada.\\Granada\\\\\\\\
\textbf{F.J. Martinez-Murcia}~\orcidlink{0000-0001-8146-7056}\\\\Dpt.
Signal Theory, Networking and Communications, University of
Granada.\\Granada\\\href{mailto:fjesusmartinez@ugr.es}{fjesusmartinez@ugr.es}\\\\\\
\textbf{I.A. Illán}\\\\Dpt. Signal Theory, Networking and
Communications, University of Granada.\\Granada\\\\\\\\
\textbf{C. Jiménez-Mesa}\\\\Dpt. Signal Theory, Networking and
Communications, University of Granada.\\Granada\\\\\\\\
\textbf{D. Castillo-Barnes}\\\\Dpt. of Communications Engineering,
University of Malaga.\\Malaga\\\\\\\\
\textbf{J. Ramírez}\\\\Dpt. Signal Theory, Networking and
Communications, University of Granada.\\Granada\\\\\\\\
\textbf{J.M. Górriz}\\\\Dpt. Signal Theory, Networking and
Communications, University of Granada.\\Granada\\}

\begin{document}
\maketitle
\begin{abstract}
This work proposes the use of 3D convolutional variational autoencoders
(CVAEs) to trace the changes and symptomatology produced by
neurodegeneration in Parkinson's disease (PD). In this work, we present
a novel approach to detect and quantify changes in dopamine transporter
(DaT) concentration and its spatial patterns using 3D CVAEs on Ioflupane
(FPCIT) imaging. Our approach leverages the power of deep learning to
learn a low-dimensional representation of the brain imaging data, which
then is linked to different symptom categories using regression
algorithms. We demonstrate the effectiveness of our approach on a
dataset of PD patients and healthy controls, and show that general
symptomatology (UPDRS) is linked to a d-dimensional decomposition via
the CVAE with R2\textgreater0.25. Our work shows the potential of
representation learning not only in early diagnosis but in understanding
neurodegeneration processes and symptomatology.
\end{abstract}
\ifdefined\Shaded\renewenvironment{Shaded}{\begin{tcolorbox}[interior hidden, boxrule=0pt, sharp corners, breakable, borderline west={3pt}{0pt}{shadecolor}, frame hidden, enhanced]}{\end{tcolorbox}}\fi

\hypertarget{sec-introduction}{%
\section{Introduction}\label{sec-introduction}}

Parkinson's disease (PD) is a progressive neurodegenerative disorder
that affects more than 6.2 million people worldwide {[}1{]}. PD is
characterized by a loss of dopamine-producing neurons in the brain,
causing tremors, stiffness, and cognitive decline among other symptoms.
FPCIT (ioflupane) SPECT is the most widely extended neuroimaging
technique for the diagnosis of PD. FPCIT binds to the presynaptic
dopamine transporters (DaTs), allowing to visualize and quantify the DaT
concentration at the striata, that characterizes PD {[}2{]}. This
modality may allow to tackle one of the biggest challenges for PD: the
early detection and monitoring of disease progression {[}3{]}.

Recent advances in deep learning, and in particular, convolutional
neural networks (CNNs) have shown promising results in medical imaging
applications such as segmentation, registration, and classification
{[}4{]}. Autoencoders (AEs), a type of self-supervised neural network,
can learn a low-dimensional representation of high-dimensional data,
making them well-suited for image compression, denoising, and anomaly
detection {[}5{]}.

In this paper, we propose a novel approach to detect and quantify subtle
changes in DaT concentration and distribution in the brain using 3D
convolutional variational AEs (CVAEs). Our approach leverages the power
of deep learning to learn a low-dimensional representation of brain
imaging data, which enables us to longitudinally compare images and
identify patterns of change that are indicative of neurodegeneration. We
demonstrate the effectiveness of our approach on a dataset of PD
patients and healthy controls, and show that latent spaces can capture
the variability of individual symptom categories, as well as the overall
disease stage.

\hypertarget{sec-materials}{%
\section{Materials and Methods}\label{sec-materials}}

\hypertarget{sec-dataset}{%
\subsection{Dataset}\label{sec-dataset}}

Data for this study were obtained from the Parkinson's Progression
Markers Initiative (PPMI) database. For the most recent information on
the study, please visit \href{}{www.ppmi-info.org}. We used the
standardized ``Original Cohort BL to Year 5'' that includes individuals
initially diagnosed as either controls (CTL) or PD affected subjects
with varying levels of severity. After removing subjects with no FPCIT
scans, the CTL group comprises 101 males and 53 females, while the PD
group comprises 284 males and 159 females, followed for up to 5 years,
with a total 1399 sessions of available data for studying the
progression of imaging biomarkers and their link to PD-specific
progression indicators. Symptomatology is assessed via the MDS-UPDRS
scale {[}6{]}, either in its aggregated form (UPDRS -total-) or its 4
parts: 1) non-motor aspects of daily living; 2) motor aspects of daily
living (tremor, walking, etc); 3) motor examination (rigidity, posture,
gait etc) and 4) motor complications (dyskinesia, fluctuation items).

FPCIT scans labeled as ``Reconstructed'' were used, having all of them a
consistent orientation and similar sizes (for a scan of shape
\(91\times 109\times 91\)). However, no spatial normalization is
performed, as it was demonstrated to have small impact when using CNN
architectures {[}4{]}. Intensity was normalized by substracting the
background average intensity, and then dividing all voxel intensities by
a non-specific reference, defined as the average voxel intensity of the
cerebellum and occipital lobe. Finally, to favor convergence of the CVAE
training, the upper-end values of the intensity distribution were
compressed using a sigmoid function.

\hypertarget{sec-vae}{%
\subsection{3D Convolutional Variational Autoencoders}\label{sec-vae}}

An autoencoder is a neural network architecture that consists of an
encoder network and a decoder network. In this work we used a 3D
convolutional encoder-decoder architecture, capable of dealing with
volumetric 3D images. The 3D encoder can capture spatial features at
multiple scales, similarly to well-known 3D convolutional NNs {[}4{]}.
Its output is a lower-dimensional representation which encodes the most
salient features of the input data. Under the manifold hypothesis
{[}5{]}, this representation is often considered as coordinates in a
``latent space'' underlying the dataset. The decoder network performs
the inverse operation.

Variational autoencoders consider that the latent space is indeed the
parameter space of \(D\) Gaussian distributions, from which the input of
the decoder is sampled {[}7{]}, allowing for non-sparse sampling and a
quantification of uncertainty in the point estimates. The \(\beta\)-VAE
loss function {[}8{]} was used (Equation~\ref{eq-vae}):

\begin{equation}\protect\hypertarget{eq-vae}{}{ \mathcal{L}_{VAE} = \mathcal{L}_{recon} + \beta \mathcal{L}_{KLD} }\label{eq-vae}\end{equation}

This loss function allows controlling for the proportion of the
reconstruction error and the Kullback-Leibler Divergence (KLD)
(Equation~\ref{eq-kld}) using the parameter \(\beta\). Reconstruction
error was estimated assuming that the voxel distribution among patients
was guassian, which means that we could use the volumetric Mean Squared
Error (MSE) (Equation~\ref{eq-recon}) between the original voxel
intensities \(x_j\) and the output voxel of the decoder \(\hat{x}_j\)
for all \(j\) voxels in the images. Each subject's loss was added to
conform the batch loss (reduction ``sum''), which proved to improve
training convergence.

\begin{equation}\protect\hypertarget{eq-recon}{}{ \mathcal{L}_{recon} = \sum_j (x_j^2 - \hat{x}_j^2)}\label{eq-recon}\end{equation}
\begin{equation}\protect\hypertarget{eq-kld}{}{ \mathcal{L}_{KLD} =  - \frac{1}{2} \sum_{i=0}^{D} \left( 1 + \log \sigma_i^2 - \mu_i^2 - \sigma_i ^2 \right) }\label{eq-kld}\end{equation}

The specific architecture and parameters of the 3D-CVAE is shown at
Table~\ref{tbl-encoder-params} and Table~\ref{tbl-decoder-params}.
Different latent dimensionality (3, 8 and 20) was tested for 400 epochs
with Adam optimizer (lr=1e-3).

\hypertarget{tbl-encoder-params}{}
\begin{longtable}[]{@{}
  >{\raggedright\arraybackslash}p{(\columnwidth - 10\tabcolsep) * \real{0.2466}}
  >{\raggedright\arraybackslash}p{(\columnwidth - 10\tabcolsep) * \real{0.1507}}
  >{\raggedright\arraybackslash}p{(\columnwidth - 10\tabcolsep) * \real{0.1507}}
  >{\raggedright\arraybackslash}p{(\columnwidth - 10\tabcolsep) * \real{0.1507}}
  >{\raggedright\arraybackslash}p{(\columnwidth - 10\tabcolsep) * \real{0.1507}}
  >{\raggedright\arraybackslash}p{(\columnwidth - 10\tabcolsep) * \real{0.1507}}@{}}
\caption{\label{tbl-encoder-params}Encoder architecture
parameters}\tabularnewline
\toprule\noalign{}
\begin{minipage}[b]{\linewidth}\raggedright
\textbf{Type}
\end{minipage} & \begin{minipage}[b]{\linewidth}\raggedright
\textbf{\# in}
\end{minipage} & \begin{minipage}[b]{\linewidth}\raggedright
\textbf{\# out}
\end{minipage} & \begin{minipage}[b]{\linewidth}\raggedright
\textbf{KS}
\end{minipage} & \begin{minipage}[b]{\linewidth}\raggedright
\textbf{Stride}
\end{minipage} & \begin{minipage}[b]{\linewidth}\raggedright
\textbf{Padding}
\end{minipage} \\
\midrule\noalign{}
\endfirsthead
\toprule\noalign{}
\begin{minipage}[b]{\linewidth}\raggedright
\textbf{Type}
\end{minipage} & \begin{minipage}[b]{\linewidth}\raggedright
\textbf{\# in}
\end{minipage} & \begin{minipage}[b]{\linewidth}\raggedright
\textbf{\# out}
\end{minipage} & \begin{minipage}[b]{\linewidth}\raggedright
\textbf{KS}
\end{minipage} & \begin{minipage}[b]{\linewidth}\raggedright
\textbf{Stride}
\end{minipage} & \begin{minipage}[b]{\linewidth}\raggedright
\textbf{Padding}
\end{minipage} \\
\midrule\noalign{}
\endhead
\bottomrule\noalign{}
\endlastfoot
Conv3D (1) & 1 & 32 & 3 & 2 & 1 \\
Conv3D (2) & 32 & 64 & 3 & 2 & 1 \\
Conv3D (3) & 64 & 128 & 3 & 2 & 1 \\
Conv3D (4) & 128 & 256 & 3 & 2 & 1 \\
Linear & 73728 & 512 & - & - & - \\
2 \(\times\) Linear (\(\mu\) and \(\sigma\)) & 512 & \(D\) & - & - &
- \\
\end{longtable}

\hypertarget{tbl-decoder-params}{}
\begin{longtable}[]{@{}
  >{\raggedright\arraybackslash}p{(\columnwidth - 10\tabcolsep) * \real{0.2727}}
  >{\raggedright\arraybackslash}p{(\columnwidth - 10\tabcolsep) * \real{0.1429}}
  >{\raggedright\arraybackslash}p{(\columnwidth - 10\tabcolsep) * \real{0.1558}}
  >{\raggedright\arraybackslash}p{(\columnwidth - 10\tabcolsep) * \real{0.1039}}
  >{\raggedright\arraybackslash}p{(\columnwidth - 10\tabcolsep) * \real{0.1558}}
  >{\raggedright\arraybackslash}p{(\columnwidth - 10\tabcolsep) * \real{0.1688}}@{}}
\caption{\label{tbl-decoder-params}Decoder architecture
parameters}\tabularnewline
\toprule\noalign{}
\begin{minipage}[b]{\linewidth}\raggedright
\textbf{Type}
\end{minipage} & \begin{minipage}[b]{\linewidth}\raggedright
\textbf{\# in}
\end{minipage} & \begin{minipage}[b]{\linewidth}\raggedright
\textbf{\# out}
\end{minipage} & \begin{minipage}[b]{\linewidth}\raggedright
\textbf{KS}
\end{minipage} & \begin{minipage}[b]{\linewidth}\raggedright
\textbf{Stride}
\end{minipage} & \begin{minipage}[b]{\linewidth}\raggedright
\textbf{Padding}
\end{minipage} \\
\midrule\noalign{}
\endfirsthead
\toprule\noalign{}
\begin{minipage}[b]{\linewidth}\raggedright
\textbf{Type}
\end{minipage} & \begin{minipage}[b]{\linewidth}\raggedright
\textbf{\# in}
\end{minipage} & \begin{minipage}[b]{\linewidth}\raggedright
\textbf{\# out}
\end{minipage} & \begin{minipage}[b]{\linewidth}\raggedright
\textbf{KS}
\end{minipage} & \begin{minipage}[b]{\linewidth}\raggedright
\textbf{Stride}
\end{minipage} & \begin{minipage}[b]{\linewidth}\raggedright
\textbf{Padding}
\end{minipage} \\
\midrule\noalign{}
\endhead
\bottomrule\noalign{}
\endlastfoot
Linear & \(D\) & 73728 & -- & -- & -- \\
ConvTranspose3d (1) & 256 & 128 & 3 & 2 & 1 \\
ConvTranspose3d (2) & 128 & 64 & 3 & 2 & 1 \\
ConvTranspose3d (3) & 64 & 32 & 3 & 2 & 1 \\
ConvTranspose3d (4) & 32 & 1 & 3 & 2 & 1 \\
\end{longtable}

\hypertarget{sec-regression}{%
\subsection{Regression}\label{sec-regression}}

The purpose of this work is to predict 4 categories of symptomatology
using the latent distribution of the FPCIT image dataset. To perform
such task, we explore the predictive power of two widely-used regression
algorithms: Decision Trees (DT) and XGBoost. DTs are relatively simple
to understand and have been widely used in medicine, because the
resulting trees are directly interpretable {[}9{]}. XGboost is a
scalable tree boosting system that uses approximate tree learning, and
is widely extended in data science, achieving high performance in many
challenges {[}10{]}.

\hypertarget{sec-experiments}{%
\subsection{Experimental Setup}\label{sec-experiments}}

The machine learning pipeline is defined by first training the dataset
and using the \(\mu\) parameters of the latent distribution as features.
Additionally, K-Means features (KMF), which capture the distance to a
K-means cluster center, were generated to capture the salient
characteristics of the data. The number of clusters used to generate
KMFs is established as the maximum value between 8 and the logarithm of
the dimensionality times 8, as in {[}11{]}. Then, the two regression
algorithms are trained with or without these KMFs to predict the
different UPDRS categories. 10-Fold cross-validation is used to compute
the performance values, including the Mean Absolute Error (MAE), Root
Mean Squared Error (RMSE) and coefficient of determination \(R^2\)
between the real and predicted values for UPDRS.

To more fully understand the output of the pipeline, we applied SHapley
Additive exPlanations (SHAP) {[}12{]} to the outputs of each system.
SHAP yields metrics such as feature importance, and the contribution of
each feature to the output of the algorithm, and thus can be used to
track the most relevant characteristics on the data.

\hypertarget{sec-results}{%
\section{Results and Discussion}\label{sec-results}}

\hypertarget{sec-regression-res}{%
\subsection{Regression Results}\label{sec-regression-res}}

\begin{figure}

{\centering \includegraphics[width=0.6\textwidth,height=\textheight]{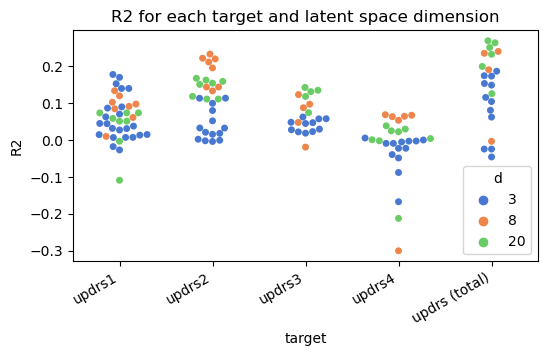}

}

\caption{\label{fig-performance}R2 performance for the models trained
with the five symptomatology scales and 3, 8 and 20 latent-dimension.}

\end{figure}

The performance of the proposed regression algorithms on the 3, 8 and
20-dimensional latent space of the CVAE is shown at
Figure~\ref{fig-performance}. We obtained R2\textgreater0 for almost all
target symptomatology categories and d-dimensional latent spaces. We can
observe that, while low latent dimensionality (3, 8) is -in general-
better for individual symptom categories, the details of the overall
symptomatology score (UPDRS -- total) is better accounted for by the
20-dimensional (20-D) CVAE, achieving a MAE = 12.21 and a R2=0.26. This
speaks of the complexity of the composite, but also of its relationship
to the FPCIT distribution patterns. Similarly, UPDRS 3, more dependent
on measured motor symptoms, benefits from the 20-D space.

A detailed view of the best results obtained for each combination of
target scale and dimensionality of the latent space D, including their
MAE, RMSE and \(R^2\) values is shown at Table~\ref{tbl-performance}.

\hypertarget{tbl-performance}{}
\begin{longtable}[]{@{}llllll@{}}
\caption{\label{tbl-performance}Classification and regression results.
DT stands for Decission Trees, XGB for XG-Boost and KMF for K-Means
features}\tabularnewline
\toprule\noalign{}
\textbf{target} & \textbf{d} & \textbf{model} & \textbf{MAE} &
\textbf{RMSE} & \textbf{R2} \\
\midrule\noalign{}
\endfirsthead
\toprule\noalign{}
\textbf{target} & \textbf{d} & \textbf{model} & \textbf{MAE} &
\textbf{RMSE} & \textbf{R2} \\
\midrule\noalign{}
\endhead
\bottomrule\noalign{}
\endlastfoot
& 3 & DT & 3.53 & 4.5 & 0.15 \\
updrs1 & 8 & DT (KMF) & 3.58 & 4.55 & 0.13 \\
& 20 & DT (KMF) & 3.69 & 4.71 & 0.07 \\
& 3 & XGB (KMF) & 3.96 & 5.23 & 0.11 \\
updrs2 & 8 & XGB (KMF) & 3.68 & 4.86 & 0.23 \\
& 20 & XGB (KMF) & 3.74 & 5.08 & 0.16 \\
& 3 & XGB (KMF) & 10.78 & 13.11 & 0.06 \\
updrs3 & 8 & XGB (KMF) & 10.21 & 12.93 & 0.09 \\
& 20 & XGB (KMF) & 10.03 & 12.59 & 0.14 \\
& 3 & DT & 0.7 & 1.44 & 0 \\
updrs4 & 8 & XGB & 0.62 & 1.39 & 0.07 \\
& 20 & DT (KMF) & 0.71 & 1.43 & 0.02 \\
& 3 & XGB (KMF) & 13.15 & 16.42 & 0.17 \\
updrs (total) & 8 & XGB & 12.72 & 15.75 & 0.24 \\
& 20 & XGB & 12.21 & 15.51 & 0.26 \\
\end{longtable}

\hypertarget{sec-interpretability}{%
\subsection{Intepretability of the Space}\label{sec-interpretability}}

The regression results show that there exist a predictive power in the
manifold representation of the FPCIT dataset. The interpretation of the
effective patterns captured by latent variables can be, in consequence,
of key importance for clinical validation. While DTs (best for UPDRS 1)
are visual by nature, ensemble methods such as XGBoost are more
difficult to interpret. Here, SHAP can pave the way to interpret the
contribution of each of the latent features to the output variables.

\begin{figure}

{\centering \includegraphics{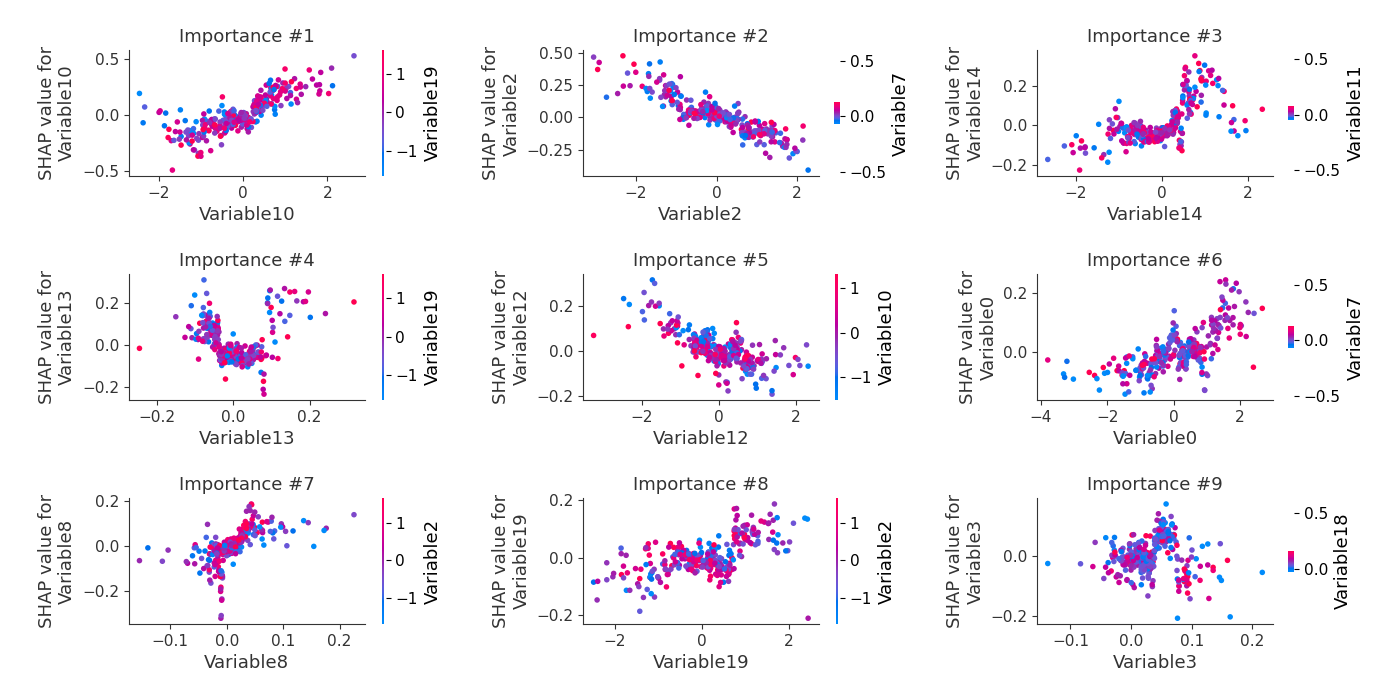}

}

\caption{\label{fig-shap-dependence}SHAP dependence plot of the XGBoost
regressor with the 20-dimensional space}

\end{figure}

In this regard, the prediction of total symptomatology (UPDRS - total)
is of special interest. In both the 8- and 20-dimensional spaces XGBoost
achieves the largest predictive ability in terms of \(R^2\).
Consequently, the most important features -measured by SHAP- can capture
the patterns that lead to this high performance. The SHAP analysis of
the XGBoost regressor with the 20-dimensional space reveals the behavior
shown at Figure~\ref{fig-shap-dependence}.

\begin{figure}

{\centering \includegraphics[width=0.54\textwidth,height=\textheight]{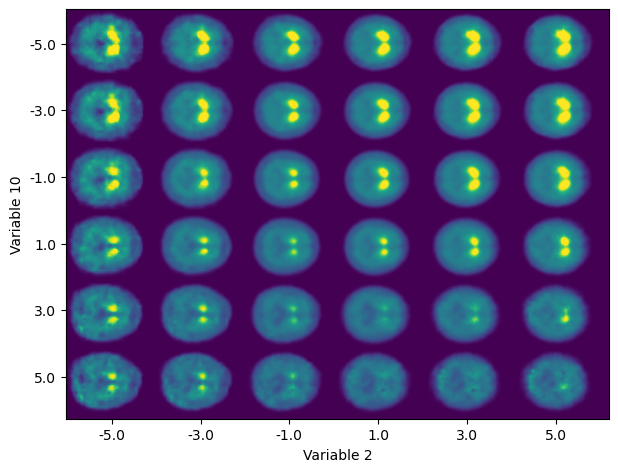}

}

\caption{\label{fig-manifold}Latent manifold for variables 10 and 2 of
the 20-dimensional latent space of the 3D CVAE.}

\end{figure}

SHAP dependence plots show how the SHAP value of importance depends on a
given feature, ordered from highest to lowest importance. In
Figure~\ref{fig-shap-dependence} we see that the top-3 features that
contribute to the output of the algorithm are variables 2, 10 and 0.
Color shows the values of a second feature that may have an interaction
effect with the feature we are plotting. Remember that these are the
mean parameters of a Gaussian distribution that encodes the pattern for
each subject. In this regard, we see an almost linear dependency between
importance to the algorithm output and values of the variables, which
indicates that there exist a clear relationship between them. To
visualize the patterns captured by features 2 and 10, we generated brain
images sampling these two features with the decoder and setting 0-mean
to all other features. The result is shown at Figure~\ref{fig-manifold}.

We observe that the composition of the two variables account for two
relevant characteristics of FPCIT: the overall intensity of the striata,
the separation between them and the ratio between uptake at the anterior
and posterior parts of the striata. General intensity and separation
patterns seem to be encoded by our variable 10, whereas the striata
anterior-posterior length is generally encoded by variable 2. In this
regard, the overall intensity (uptake) at the striata is indeed the
objective of FPCIT as a biomarker, and therefore expected. As for the
anterior-posterior striata patterns, many works that used FDOPA, a
presynaptic dopaminergic marker, reported a distinct anterior--posterior
gradient of uptake as PD progresses {[}13{]}. This could be then a
relevant marker for assessing the progression of the disease. Since SHAP
importance is higher for this variable 2, we can assume that its
contribution to the modelling of the progression is higher, perhaps
showing a more smooth transition as the symptoms progress, in contrast
to overall drug uptake, more related to late neurodegeneration.

\hypertarget{sec-conclusions}{%
\section{Conclusions}\label{sec-conclusions}}

We hypothesize that the non-linear self-supervised latent space of a 3D
Convolutional Variational Autoencoder (CVAE) is linked to the symptoms
shown by affected subjects. The latent features of a trained CVAEs was
related to different aspects of the MDS-UPDRS scale with
R2\textgreater0.20, proving a link between FPCIT spatial patterns and
Parkinson's Disease symptomatology. The most relevant variables for the
predictive algorithms reveal that the CVAE captures patterns related to
overall intensity, striata separation and differences between
anterior-posterior parts of the striata, and that this last feature is
more relevant for assessing the progression of symptoms than the former.
This comprehensive approach may help to better understand the complex
relationship between clinical disease and imaging biomarkers, beyond the
obvious relationship between the average binding-ratio and the disease.

\hypertarget{acknowledgements}{%
\section*{Acknowledgements}\label{acknowledgements}}
\addcontentsline{toc}{section}{Acknowledgements}

This work was partly supported by the MINECO/FEDER under the
RTI2018-098913-B-I00 projects, and in by the Consejería de Economía,
Innovación, Ciencia y Empleo (Junta de Andalucía) and FEDER under the
P20-00525 project and PPJIA2021-17. Work by F.J.M.M. was supported by
the MICINN ``Ramón y Cajal'' RYC2021-030875-I. Work by C.J.M. is
supported by Ministerio de Universidades under the FPU18/04902 grant.

\hypertarget{references}{%
\section*{References}\label{references}}
\addcontentsline{toc}{section}{References}

\hypertarget{refs}{}
\begin{CSLReferences}{0}{0}
\leavevmode\vadjust pre{\hypertarget{ref-feigin2017global}{}}%
\CSLLeftMargin{{[}1{]} }%
\CSLRightInline{V.L. Feigin, A.A. Abajobir, K.H. Abate, et al.,
{``Global, regional, and national burden of neurological disorders
during 1990--2015: A systematic analysis for the global burden of
disease study 2015,''}. \emph{The Lancet Neurology}. vol. 16, no. 11,
pp. 877--897, 2017.}

\leavevmode\vadjust pre{\hypertarget{ref-segoviaPreprocessing18FDMFPPETData2017}{}}%
\CSLLeftMargin{{[}2{]} }%
\CSLRightInline{F. Segovia, J.M. Górriz, J. Ramírez, F.J.
Martínez-Murcia, and D. Salas-Gonzalez, {``Preprocessing of
{18F}-{DMFP}-{PET Data Based} on {Hidden Markov Random Fields} and the
{Gaussian Distribution},''}. \emph{Frontiers in aging neuroscience}.
vol. 9, p. 326, 2017.}

\leavevmode\vadjust pre{\hypertarget{ref-tolosa2021challenges}{}}%
\CSLLeftMargin{{[}3{]} }%
\CSLRightInline{E. Tolosa, A. Garrido, S.W. Scholz, and W. Poewe,
{``Challenges in the diagnosis of parkinson's disease,''}. \emph{The
Lancet Neurology}. vol. 20, no. 5, pp. 385--397, 2021.}

\leavevmode\vadjust pre{\hypertarget{ref-martinez-murciaConvolutionalNeuralNetworks2018}{}}%
\CSLLeftMargin{{[}4{]} }%
\CSLRightInline{F.J. Martinez-Murcia, J.M. Górriz, J. Ramírez, and A.
Ortiz, {``Convolutional neural networks for neuroimaging in parkinson's
disease: Is preprocessing needed?''}. \emph{International journal of
neural systems}. vol. 28, no. 10, p. 1850035, 2018.}

\leavevmode\vadjust pre{\hypertarget{ref-martinez-murciaStudyingManifoldStructure2020}{}}%
\CSLLeftMargin{{[}5{]} }%
\CSLRightInline{F.J. Martinez-Murcia, A. Ortiz, J.-M. Gorriz, J.
Ramirez, and D. Castillo-Barnes,
{``\href{https://doi.org/10.1109/JBHI.2019.2914970}{Studying the
{Manifold Structure} of {Alzheimer}'s {Disease}: {A Deep Learning
Approach Using Convolutional Autoencoders}},''}. \emph{IEEE Journal of
Biomedical and Health Informatics}. vol. 24, no. 1, pp. 17--26, 2020.}

\leavevmode\vadjust pre{\hypertarget{ref-goetz2008movement}{}}%
\CSLLeftMargin{{[}6{]} }%
\CSLRightInline{C.G. Goetz, B.C. Tilley, S.R. Shaftman, et al.,
{``Movement disorder society-sponsored revision of the unified
parkinson's disease rating scale (MDS-UPDRS): Scale presentation and
clinimetric testing results,''}. \emph{Movement disorders: official
journal of the Movement Disorder Society}. vol. 23, no. 15, pp.
2129--2170, 2008.}

\leavevmode\vadjust pre{\hypertarget{ref-kingma2013auto}{}}%
\CSLLeftMargin{{[}7{]} }%
\CSLRightInline{D.P. Kingma and M. Welling, {``Auto-encoding variational
bayes,''}. \emph{arXiv preprint arXiv:1312.6114}. 2013.}

\leavevmode\vadjust pre{\hypertarget{ref-higgins2017beta}{}}%
\CSLLeftMargin{{[}8{]} }%
\CSLRightInline{I. Higgins, L. Matthey, A. Pal, et al., {``Beta-vae:
Learning basic visual concepts with a constrained variational
framework,''}. In: \emph{International conference on learning
representations} (2017).}

\leavevmode\vadjust pre{\hypertarget{ref-arco2023uncertainty}{}}%
\CSLLeftMargin{{[}9{]} }%
\CSLRightInline{J.E. Arco, A. Ortiz, J. Ramírez, F.J. Martínez-Murcia,
Y.-D. Zhang, and J.M. Górriz, {``Uncertainty-driven ensembles of
multi-scale deep architectures for image classification,''}.
\emph{Information Fusion}. vol. 89, pp. 53--65, 2023.}

\leavevmode\vadjust pre{\hypertarget{ref-chen2016xgboost}{}}%
\CSLLeftMargin{{[}10{]} }%
\CSLRightInline{T. Chen and C. Guestrin, {``Xgboost: A scalable tree
boosting system,''}. In: \emph{Proceedings of the 22nd acm sigkdd
international conference on knowledge discovery and data mining}. pp.
785--794 (2016).}

\leavevmode\vadjust pre{\hypertarget{ref-boutsidis2010random}{}}%
\CSLLeftMargin{{[}11{]} }%
\CSLRightInline{C. Boutsidis, A. Zouzias, and P. Drineas, {``Random
projections for \(k\)-means clustering,''}. \emph{Advances in neural
information processing systems}. vol. 23, 2010.}

\leavevmode\vadjust pre{\hypertarget{ref-SHAP2017}{}}%
\CSLLeftMargin{{[}12{]} }%
\CSLRightInline{S.M. Lundberg and S.-I. Lee,
{``\href{https://proceedings.neurips.cc/paper/2017/file/8a20a8621978632d76c43dfd28b67767-Paper.pdf}{A
unified approach to interpreting model predictions},''}. In: I. Guyon,
U.V. Luxburg, S. Bengio, et al., Eds. \emph{Advances in neural
information processing systems}. \emph{Curran Associates, Inc.} (2017).}

\leavevmode\vadjust pre{\hypertarget{ref-dhawan2002comparative}{}}%
\CSLLeftMargin{{[}13{]} }%
\CSLRightInline{V. Dhawan, Y. Ma, V. Pillai, et al., {``Comparative
analysis of striatal FDOPA uptake in parkinson's disease: Ratio method
versus graphical approach,''}. \emph{Journal of Nuclear Medicine}. vol.
43, no. 10, pp. 1324--1330, 2002.}

\end{CSLReferences}

\end{document}